# An Interactive System for Exhibitions in a Science and Technology Center

Alessio De Angelis, Paolo Carbone, Marco Dionigi,
Emilio Di Giacomo, Aurelio Stoppini, Fabio
Radicioni
Department of Engineering
University of Perugia, Perugia, Italy,
E-mail: alessio.deangelis@unipg.it

Enrico Tombesi
Perugia Officina Scienza Tecnologia (POST)
Science and Technology Center
Perugia, Italy

*Abstract*—**This paper presents the development of a system for realizing interactive exhibitions in the context of a science and technology center. The core functionality of the system is provided by a positioning subsystem comprised of a fixed infrastructure of transmitters and a sensor worn by a user. The operating principle of the positioning system is based on inductive coupling of resonators. Information about the position of the user is transferred to an information system for processing and displaying. Possible use cases include interactive games, information retrieval interfaces and educational scenarios.**

*Keywords—Science and technology centers, Interactive exhibitions, Indoor positioning.*

## I. Introduction

Science and technology centers play a fundamental role for promoting the education of the general public on scientific topics, which are pervasive in today's society. To stimulate the interest of a diverse audience, a science and technology center typically provides interactive exhibitions. In this context, several different electronic and information systems have been described in the literature, ranging from the immersive augmented reality system presented in [1], to pervasive social games proposed for cultural organizations in [2]. Additionally, traditional museum exhibitions are benefitting from the use of technology, such as decision support systems, to plan exhibitions for an optimized user experience. This aspect is enabled by positioning systems deployed inside the exhibition center to track visitors and analyze their behavior, such as those presented in [3] and [4].

In this paper, the development of an interactive system for engaging and educating visitors of a science and technology center is described. This system employs position and movement in two dimensions as a means for user interaction with an exhibition. Therefore, the core of the realized system is a positioning subsystem based on an infrastructure of known-position nodes deployed in the exhibition area and a sensor, which is worn by the user. The principle of operation of the positioning system is based on inductive coupling of resonating coils [5], [6]. The position measurement results are used by a software application to implement the interaction functionality and game mechanics of the interactive exhibition. Such software application also provides the information feedback to the user by means of a display installed in the exhibition.

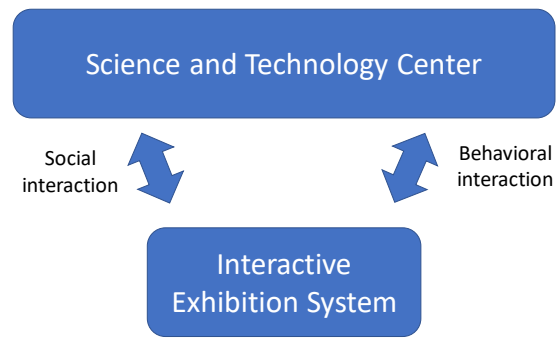

Fig. 1 – Interactions of the developed system with the larger system of which it is a part, i.e. the Science and Technology Center.

Potentially, the educational value of the interactive system lies both in the application and in the principle of operation of the positioning system itself. On the one hand, by properly designing the software application as a game, it is possible to demonstrate scientific concepts in an entertaining fashion. On the other hand, by providing additional information material, it is possible to educate the visitor on the underlying scientific concepts of electromagnetism and geometric localization methods, while the visitor is experiencing the effects of those concepts.

## II. System Interactions and Design Requirements

The science and technology center is a system of systems and therefore the interactions between the developed interactive exhibition system and the larger system represented by the science and technology center must be analyzed. Such interactions include social, behavioral, and reliability aspects, as illustrated in Fig. 1.

The social interaction aspect is related to the kind of audience the system is targeted to, within the people who visit the science and technology center, namely, children, teenagers, and adults. To stimulate the interests of these different types of audience, the system should support several different activities. In the developed system, this is accomplished by a software application, called *Position Controlled Application*, that offers various activities/games and a selection interface where the user can select the most relevant game.

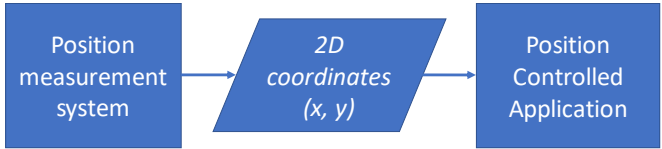

Fig. 2 – Architecture of the realized interactive exhibition system.

The behavioral interaction aspect relates to the type of behavior that the system requires from its users, e.g. the kind of motion and gestures that the user performs, and the ergonomics implications on the design. In the developed system, the ergonomic aspect was considered during the design phase and, as a result, a wearable device was realized, which the user can wear as a badge during the walking motion or hold in the hand. Finally, it is important to consider reliability aspects such as the possible stress and damage caused to the realized system due to misuse, improper handling of hardware components, or incidents by the exhibition visitors.

The application, together with the system interactions described above, dictates several constraints that must be considered during the design phase. In particular, the developed system should be intuitive and easy to use, physically robust, self-contained, and responsive. These constraints may be translated into technical requirements, to guide the design process. Specifically, the designed system should not require user configuration of hardware and software components, it should be easily wearable or held by the user without cabling or power cords, and finally it should provide position information with a decimeter-order accuracy and an update rate of at least 1 Hz. The developed system, which is described and characterized in the following sections, complies with these requirements.

## III. System Architecture

### A. High-level architecture

The architecture of the realized system is shown in Fig. 2. The two coordinates $(x, y)$, representing the position of the user in a coordinate system relative to the known node infrastructure, are continuously measured by the position measurement system and provided as an input to the Position Controlled Application. This application processes the position according to the game mechanics and provides feedback to the user by displaying the information graphically on a monitor.

### B. Position measurement subsystem architecture

The architecture of the position measurement subsystem is shown in Fig. 3. An infrastructure consisting of four known-position transmitting coils (*anchors*) is deployed. Each anchor transmits a sinusoidal signal and is characterized by a unique operating frequency; thus, the system operates in frequency division multiplexing mode [5]. The user wears a receiving coil mounted on a badge (*mobile node*), which is equipped with a microcontroller that digitizes the received signal and transfers the samples to the PC via a Bluetooth connection. The mobile node is self-contained, battery operated, and enclosed in a 3D-printed case measuring approximately $15 \times 5 \times 5$ cm, which may be attached to the user as a badge.

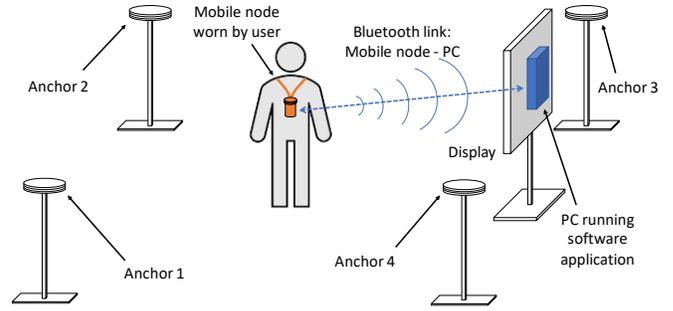

Fig. 3 – Architecture of the positioning subsystem.

Acquired samples are processed by the PC to estimate the amplitude of the signals received by the four transmitting anchors. The amplitude estimation algorithm is based on the linear least squares sinefit procedure in [7], extended to the case of a signal consisting of the sum of four sinusoids. The estimated amplitude values are then used to compute the distance between each anchor and the mobile node by inverting the model that relates induced voltage $V$ at the receiver to transmitter-receiver distance $d$.

Under the assumption that the receiver and transmitter coils lie on the same plane, it may be shown that such model is represented by a power law, i.e.

$$V = \alpha d^{-\beta}, \qquad (1)$$

where $\alpha$ and $\beta$ are constants, different for each anchor, and dependent on component tolerances and environmental characteristics [6]. The ideal free-space value of $\beta$ is 3. However, if the environment surrounding the coils contains metallic structures, the magnetic field might be distorted and such value might differ from the ideal value. In practical applications, the values of $\alpha$ and $\beta$ should be found by calibration in the preliminary deployment phase of the system.

The power-law model in (1) allows for a low-complexity processing method to be implemented, and for realizing a low-power 2-D positioning system. If the assumption that the receiver and transmitter coils lie on the same plane is not applicable, i.e. for arbitrary orientations, a different model must be used, which is more computationally burdensome and prone to numerical issues [5]. In the developed system, we apply the simple power-law model in (1), due to its computational advantages. Therefore, we design the hardware configuration of the system to ensure that all coils lie approximately on the same plane. In particular, the anchor coils are mounted horizontally on stands at the same height and the receiving coil is mounted horizontally on the mobile node carried by the user. Small deviations from this configuration are unavoidable in practice, mainly due to user height difference, posture, and walking motion. However, their influence on positioning accuracy is within acceptable bounds. As shown in [5], in fact, negligible positioning errors occur if the transmitter and receiver coils do not lie on the same plane, provided that the angle between the line passing through the centers of a transmitter and the receiver and the plane of the transmitters is smaller than 10°.

Once distance estimates are available, the two-dimensional vector containing the *x* and *y* coordinates of the user's position, denoted by $\mathbf{x} \equiv [x\ y]^T$, is estimated by trilateration, i.e. by solving a least squares optimization problem as follows [5]:

$$\hat{\mathbf{x}} = \arg\min_{\mathbf{x}} \sum_{i=1}^{N} (\tilde{d}_i - \|\mathbf{x} - \mathbf{a}_i\|)^2 \qquad (2)$$

where $\tilde{d}_i$ is the measured distance between the mobile node and the *i*-th anchor, *N* is the number of anchors, with $N = 4$ in the realized system, $\mathbf{a}_i$ is the vector of the known coordinates of the *i*-th anchor, and $\|\cdot\|$ denotes the Euclidean norm.

*C. Architecture of the Position Controlled Application*

As explained in Section II, the position of the user continuously measured by the position measurement system is provided as an input to a software application, called Position Controlled Application (PCA). PCA offers various activities/games (we will refer to them as *apps*) that can be executed/played by the users. The architecture of PCA is shown in Fig. 4. The *Position Receiver* module is a server listening on a socket; the positioning subsystem acts as a client for this server continuously sending the measured coordinates. The received values are forwarded to the *Execution Environment* module. This communication is realized according to the *observer* design pattern: each time a pair of coordinates is received, the status of a suitable object is changed and the *Execution Environment* module (the observer) is notified. As the name suggests, the *Execution Environment* module represents the "environment" that allows the various apps to be executed. It has three main functionalities:
1. It maintains a list of available apps. This list contains not only the activities/games to be played by the user, but also some utility apps. In each moment, one of the apps in the list is in execution. When the application is launched the app in execution is the *Home App*, a utility app that presents to the user the list of available apps, and allows her to select the one she is interested in.
2. It maps the physical coordinates received by the position measurement system into pixel coordinates in the application canvas. The received coordinates (*x*,*y*) indicate a position in the physical space where the user can move; in order for these coordinates to be used by the apps they need to be translated into a position (*x'*,*y'*) in the virtual space of the application canvas. The transformation is given by:

$$x' = \frac{W}{(x_{MAX} - x_{MIN})}(x - x_{MIN})$$

$$y' = \frac{H}{(y_{MAX} - y_{MIN})}(y - y_{MIN})$$

where *W* and *H* are the width and the height of the canvas, respectively, while $x_{MAX}$ and $x_{MIN}$ ( $y_{MAX}$ and $y_{MIN}$, respectively) indicate the maximum and minimum *x*-coordinate (*y*-coordinate, respectively) in the physical space. In order to know the values of $x_{MAX}$, $x_{MIN}$, $y_{MAX}$ and $y_{MIN}$, a *Calibration App* is available. This app performs a simple calibration procedure asking the user to move to the four sides of the space where she can move and registering the values received by the position measurement system.

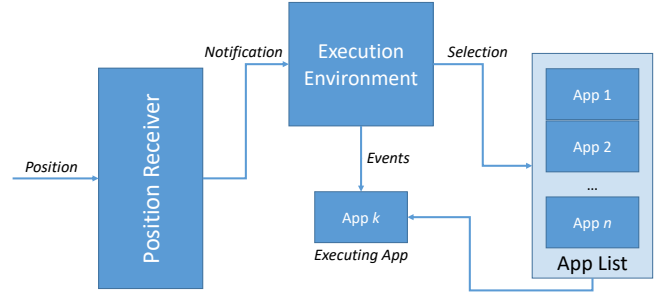

Fig. 4 – The PCA architecture.

3. It generates events for the app that is executing. Once the coordinates are transformed, the *Execution Environment* module generates an event of type *UserMoved*, thus indicating to the executing app that the user has moved to position (*x'*,*y'*). Each app handles the event according to its logic. Furthermore, if the position received remains the same for a certain number of times (five in the current implementation) a *UserClicked* event is generated. This event is meant to be used by the apps as the equivalent of a mouse click.

The application has been implemented by using the JavaFX 8 technology.

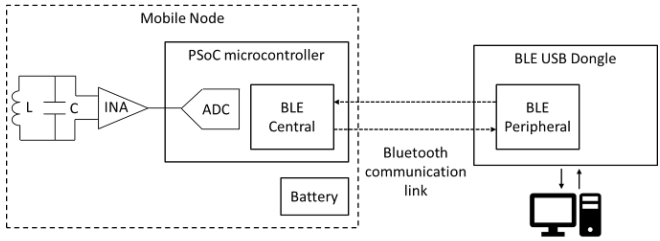

Fig. 5 - Architecture of the mobile node (receiver), showing its wireless connection with the PC.

## IV. POSITIONING SUBSYSTEM IMPLEMENTATION

### A. Implementation Issues

The transmitting anchor nodes are implemented as air-core copper-wire coils having a radius of 7 cm and 20 windings, resulting in a nominal inductance value of approximately 125 µH. A 160 nF capacitor is connected in to each coil, thus realizing a parallel LC resonator. The resulting nominal resonance frequency is approximately 35.6 kHz, and the quality factor of the resonator is on the order of 10. The receiving mobile node is realized using a smaller coil, since it must be wearable and easily carried by the user. The radius of the receiving coil is approximately 2.5 cm, and the coil has 60 turns, resulting in a nominal inductance of approximately 200 µH. A 100 nF capacitor is connected to the receiving coil, to obtain the same nominal resonance frequency as that of the transmitters. The total power consumption of each anchor is approximately 100 mW, and each anchor is supplied by a wall power supply.

Each anchor is programmed with a unique operating frequency, which is close to the resonance of the circuit. A quartz oscillator is used at each anchor to stabilize the operating frequency. The set of frequencies assigned to the four anchors is the following: [34482.7 35398.2 36144.5 36922.8] Hz. The coil is driven by a device based on a programmable system on chip (PSoC) microcontroller, of the PSOC 5 LP family, which generates a 5-Vpp square wave at the specified frequency. Moreover, a transistor driver circuit provides the required current to the coil, and the resonance behavior results in a sinusoidal time-varying magnetic field.

A diagram depicting the architecture of the realized mobile node, which acts as receiver, is shown in Fig. 5. The voltage signal induced by the time-varying magnetic field generated by the anchors, consisting of the sum of four sinusoidal signals, is first amplified using an instrumentation amplifier with a gain of 100. Then, it is digitized by the analog-to-digital converter included in a microcontroller of the PSoC 4 Bluetooth Low Energy (BLE) family, sampling at 200 kSa/s, 12 bits. A record of 300 samples is acquired and transferred to the PC using the Bluetooth link for further processing. This procedure is repeated as soon as the BLE data transmission is completed. The Bluetooth scan, handshaking, and connection operations are performed in the initial deployment phase of the system, thus without requiring user intervention.

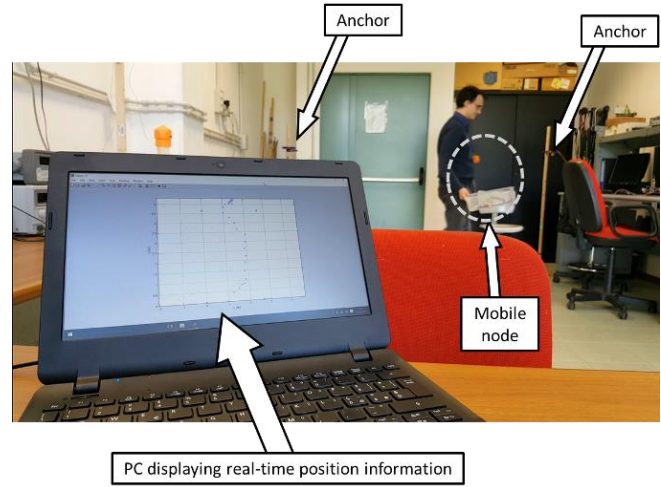

Fig. 6 - Picture of the preliminary experimental setup.

### B. Preliminary test setup

A picture of the preliminary setup used for developing and characterizing the system in controlled laboratory conditions is shown in Fig. 6. The anchors were placed at a height of approximately 1.2 m on the corners of a 2.66 m × 4.66 m rectangle. The user carried the receiving coil as a badge, the receiver's electronics and battery were contained in a box carried by the user, for ease of access and modification. This setup allowed for rapid functional verification of the system operation. Before operating the system, a calibration procedure was performed, to identify the $\alpha$ and $\beta$ constants in (1) for each anchor. The calibration was performed by placing the receiver at five calibration points having known distances from each anchor and calculating the values of $\alpha$ and $\beta$ in (1) that best fit the measured values in a least squares sense. In the system's operating phase, a position update rate of 2 Hz was observed, and the system could consistently estimate the trajectory of a user during walking motion in real time, inside the area delimited by the four anchors, as can be seen in Fig. 6.

### C. Positioning accuracy evaluation

To evaluate the positioning accuracy, a geodetic survey has been performed, obtaining the reference positions. By means of an electro-optical geodetic theodolite (Leica Geosystems total station TS-06) the three-dimensional coordinates of the following points have been determined:

- centers of the four transmitting coils A, B, C, D realizing the datum;

- center of the receiving test solenoid occupying the calibration points C1, C2, C3, C4, C5;

- center of the receiving test solenoid occupying the control points P01…P27.

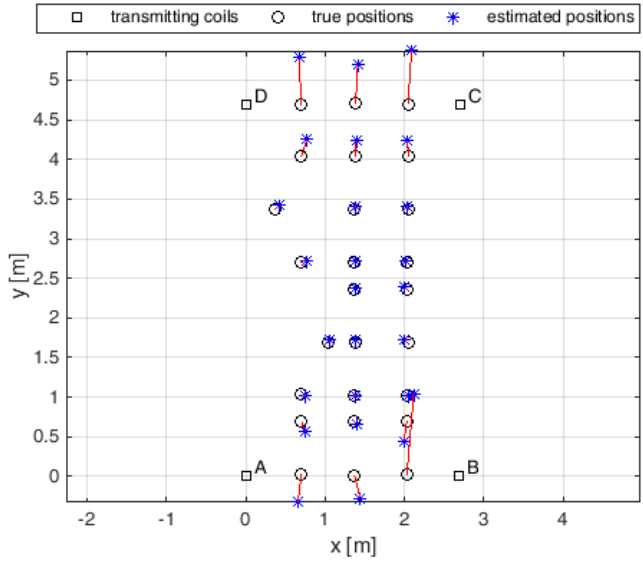

Fig. 7 – Experimental positioning results. Average estimated positions are shown as asterisks for each control point. Relative positions are in the coordinate frame centered at A in (0,0).

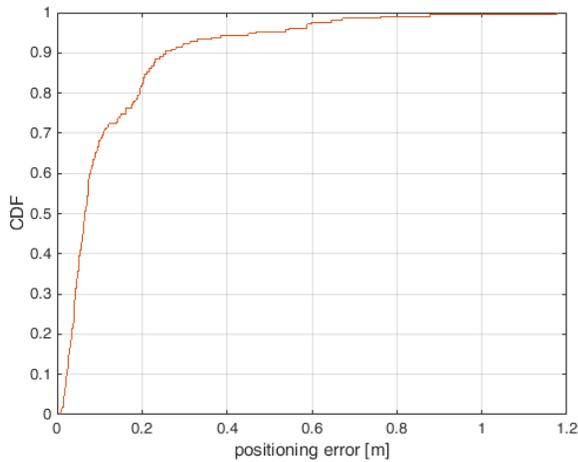

Fig. 8 – Empirical CDF of the positioning error.

TABLE I – COORDINATES IN THE LOCAL SYSTEM FROM THE GEODETIC SURVEY

| Points | x (m) | y (m) | z (m) |
|---|---|---|---|
| A | 0.000 | 0.000 | 1.250 |
| B | 2.678 | 0.000 | 1.263 |
| C | 2.711 | 4.694 | 1.242 |
| D | 0.009 | 4.692 | 1.233 |
| C1 | 0.700 | 1.018 | 1.237 |
| C2 | 2.037 | 1.023 | 1.240 |
| C3 | 2.044 | 3.699 | 1.235 |
| C4 | 0.701 | 3.703 | 1.230 |
| C5 | 1.367 | 2.360 | 1.235 |
| P01 | 0.699 | 0.023 | 1.240 |
| P02 | 1.371 | 0.006 | 1.240 |
| P03 | 2.030 | 0.016 | 1.242 |
| P04 | 0.703 | 0.685 | 1.239 |
| P05 | 1.357 | 0.703 | 1.237 |
| P06 | 2.030 | 0.695 | 1.240 |
| P07 | 0.701 | 1.029 | 1.239 |
| P08 | 1.371 | 1.021 | 1.240 |
| P09 | 2.035 | 1.019 | 1.240 |
| P10 | 1.039 | 1.694 | 1.237 |
| P11 | 1.374 | 1.690 | 1.237 |
| P12 | 2.044 | 1.686 | 1.239 |
| P14 | 1.368 | 2.365 | 1.236 |
| P15 | 2.041 | 2.360 | 1.238 |
| P16 | 0.697 | 2.703 | 1.233 |
| P17 | 1.371 | 2.699 | 1.236 |
| P18 | 2.041 | 2.701 | 1.238 |
| P19 | 0.371 | 3.370 | 1.230 |
| P20 | 1.370 | 3.368 | 1.234 |
| P21 | 2.044 | 3.365 | 1.235 |
| P22 | 0.703 | 4.037 | 1.229 |
| P23 | 1.378 | 4.039 | 1.233 |
| P24 | 2.052 | 4.046 | 1.232 |
| P25 | 0.697 | 4.698 | 1.228 |
| P26 | 1.384 | 4.710 | 1.231 |
| P27 | 2.045 | 4.696 | 1.233 |
| A* | 0.000 | 0.002 | 1.250 |
| B* | 2.677 | 0.001 | 1.263 |
| C* | 2.710 | 4.694 | 1.243 |
| D* | 0.009 | 4.690 | 1.234 |

To perform the measurements, the datum solenoids and the test solenoid have been signalized by means of square retro-reflective targets (58x58 mm size) positioned over the vertical axis of each solenoid at a vertical offset measured at a ±1 mm accuracy. The accuracy obtained by the TS-06 on the described targets in the experimental conditions can be prudently assumed as ±2 mm. For control purposes, all geodetic measures have been repeated five times, and the datum points have been measured twice, at the start and at the end of the survey, for a repeatability check which showed differences less than 2 millimeters. The results of the geodetic survey are resumed in Table I, where A*, B*, C*, and D* denote the coordinates obtained by measuring the datum points the second time, at the end of the survey.

After the geodetic survey, 10 repeated position measurements were performed using the realized positioning system at each control point. The positioning error was defined as the Euclidean distance between the 2D position estimated by the proposed system and the 2D reference position, obtained by projecting the position provided by the geodetic survey on the plane where the transmitting coil A lies.

Experimental results are shown in Fig. 7, where it can be noticed that the positioning error in the central part of the area delimited by the anchors is smaller than in the border regions. This is mainly due to unfavorable geometric configurations that result in a poor geometrical dilution of precision [9], and to saturation effects when the receiver is too close to a transmitter. By considering all surveyed positions, the mean positioning error was 21.5 cm. By eliminating the six control points on the borders, thus considering only the control points that are strictly inside the region delimited by the transmitting nodes, the mean positioning error was 12.4 cm. Therefore, the decimeter-order accuracy requirement is satisfied by the realized system. Furthermore, the empirical cumulative distribution function (CDF) of the positioning error for the control points inside the region delimited by the transmitting nodes is shown in Fig. 8. It is possible to notice that, in 90% of the cases, the error is smaller than 25 cm. The accuracy and coverage area of the system could be further improved by using additional anchors. However, the obtained levels of accuracy and coverage are satisfactory for the application under consideration.

## V. CONCLUSION

In this paper, an indoor positioning system was presented and characterized. The aim of this system is to enable the development of interactive exhibitions in a science and technology center. The architecture of the developed system, comprised of a position measurement subsystem and a position controlled application, was presented. Experimental results were provided for position accuracy characterization, showing a decimeter-order error.

### ACKNOWLEDGMENTS

This research activity was funded through the INCOMPASS project, Fondo Ricerca di Base 2016, by the University of Perugia and through grant PRIN 2015C37B25 by the Italian Ministry of Instruction, University and Research (MIUR), whose support the authors gratefully acknowledge.